\documentclass[slac_one]{revtex4}
\usepackage{graphicx}
\usepackage{fancyhdr}
\begin{document}

\title{Hadronic Tau Decays at Belle} 

%

\author{K. Inami}
\affiliation{Nagoya university, Nagoya, 464-8602, JAPAN}
\author{K. Hayasaka}
\affiliation{Nagoya university, Nagoya, 464-8602, JAPAN}
\author{M. J. Lee}
\affiliation{Seoul national university, Seoul, 151-742, KOREA}
\author{Y. Usuki}
\affiliation{Nagoya university, Nagoya, 464-8602, JAPAN}
\author{and Belle collaboration}

\begin{abstract}
We present a study of hadronic tau decays, especially decay into
three-hadron final states with kaon(s),
using more than $500$fb$^{-1}$ of data collected with the Belle detector 
at the KEKB asymmetric-energy $e^+e^-$ collider.
We measure the branching ratios both inclusively
and exclusively while considering the intermediated states of $K^*(892)^0$.
The following results are obtained:
${\cal B}(\tau^- \to K^- \pi^+ \pi^- \nu_\tau)=
(3.25 \pm 0.02^{+0.16}_{-0.15})\times 10^{-3}$,
${\cal B}(\tau^- \to K^- K^+ \pi^- \nu_\tau)=
(1.53 \pm 0.01\pm 0.06)\times 10^{-3}$,
${\cal B}(\tau^- \to K^- K^+ K^- \nu_\tau)=
(2.60 \pm 0.23\pm 0.10)\times 10^{-5}$
and
${\cal B}(\tau^-\to K^*(892)^0 K^-\nu_\tau)=(1.56\pm0.02\pm0.09)\times10^{-3}$.
We also measure the mass and width of the $K^*(892)^0$ resonance as
$M_{K^*(892)^0} = 895.10\pm0.27\pm0.31$~MeV/$c^2$ and
${\mit\Gamma}_{K^*(892)^0} = 47.23\pm0.49\pm0.79$~MeV,
which are the first results obtained using $\tau$ decay.
\end{abstract}

\maketitle

\thispagestyle{fancy}



\section{Introduction}
Hadronic $\tau$ decays with kaons provide a good probe for the strange sector
of a weak charged current.
The decays of $\tau^- \to K^- \pi^+ \pi^- \nu_\tau$ and
$\tau^- \to K^- K^+ K^- \nu_\tau$ are Cabibbo suppressed decay.
From the study of these modes, one can determine the strange quark mass
and the CKM matrix of $V_{us}$.
On the other hand,
the decay of $\tau^- \to K^- K^+ \pi^- \nu_\tau$,
including $\tau^- \to K^{*0} K^- \nu_\tau$ occurs through a
vector and axial-vector current. Therefore, it is sensitive to
the Wess-Zumino anomaly for SU(3)$_L \times $SU(3)$_R$.

This study was performed at the KEKB asymmetric-energy $e^+e^-$ 
collider~\cite{KEKB} with the Belle detector~\cite{BelleDet}.
The KEKB collider has realized the world-highest luminosity of
$1.7 \times 10^{34}$cm$^{-2}$s$^{-1}$.
The Belle detector shows good particle identification (PID)
ability. PID is based on the energy deposit and shower shape in the 
electromagnetic calorimeter, 
the momentum and $dE/dX$ measured in the drift chamber, the particle range 
in the muon chamber, the light yield in the aerogel threshold
Cherenkov counters, and the particle's time-of-flight
from the TOF counter~\cite{EMu}.
The efficiency and fake rate for kaon identification is
about 90\% and 6\%, respectively.

In following studies,
the detection efficiency for each signal mode and the amount of 
the background (BG) contribution are estimated from the data itself
as well as Monte Carlo (MC) simulations. 
To generate signal events as well as $\tau$-decay 
originated background, the KKMC program~\cite{KKMC} is used. The background 
from the $e^+e^-\to q\bar{q}$ process is simulated 
using EvtGen~\cite{EvtGen}.
The detector response is simulated by a GEANT3~\cite{GEANT} based program.

\section{Study of $\tau^- \to K^- \pi^+ \pi^- \nu_\tau$}

In the analysis, we select 3-1 topology events divided by the thrust
axis, and require no energetic photon on the 3-prong signal side.
The large amount of $q\bar{q}$ and two-photon BG processes are 
removed by requiring the missing energy to be due to the $\tau$ neutrinos
in the signal candidates, and the lepton on the tag side.
We apply the kaon identification to the charged particles on the signal side
and categorize the final state. The main BG source after event selection
is the $\tau$ decays with a mis-identified pion as a kaon.
Therefore, 
in order to suppress the systematic uncertainty,
we analyze the cross-feed BG events simultaneously.

\begin{figure}[t]
\centering
\includegraphics[width=17cm]{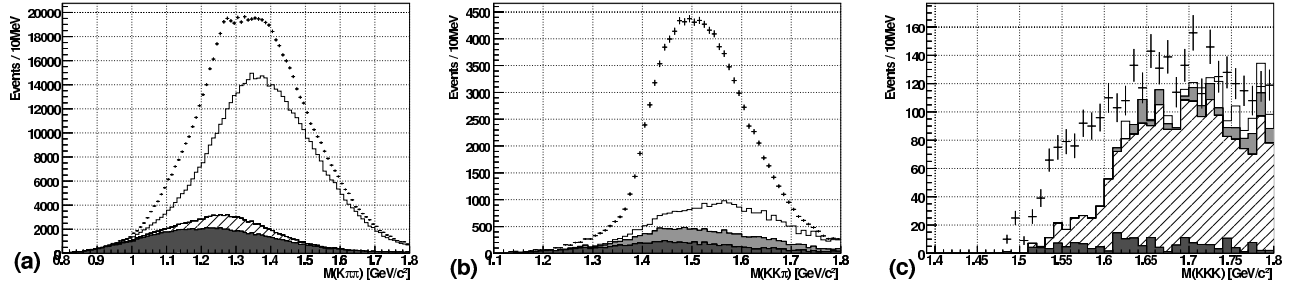}
\caption{Invariant mass distributions of
(a) $K^- \pi^+ \pi^-$, (b) $K^- K^+ \pi^-$ and (c) $K^- K^+ K^-$.
Cross points are data.
Histograms indicate the BG distributions estimated by MC
for $\tau^- \to \pi^-\pi^+\pi^-\nu_\tau$ (open),
$\tau^- \to K^-\pi^+\pi^-\nu_\tau$ (light gray),
$\tau^- \to K^-K^+\pi^-\nu_\tau$ (hatched)
and other BG modes (dark gray).
}
\label{fig:m3h}
\end{figure}

Figures~\ref{fig:m3h} show the invariant mass distributions of
$K^- \pi^+ \pi^-$, $K^- K^+ \pi^-$ and $K^- K^+ K^-$ for samples
with the corresponding PID selection, using data of 669fb$^{-1}$.
The contribution of $q\bar{q}$ and two-photon processes are found to be
negligible from a MC estimation.
The decays of $\tau^- \to K^- \pi^+ \pi^- \nu_\tau$ and 
$\tau^- \to K^- K^+ K^- \nu_\tau$
have large BG contributions due to a mis-identification of kaons.
On the other hand, the result of $\tau^- \to K^- K^+ \pi^- \nu_\tau$ 
shows less BG because of a relatively high branching fraction.

Considering the cross-feed contributions between
$\tau^- \to \pi^-\pi^+\pi^-\nu_\tau$, $K^-\pi^+\pi^-\nu_\tau$,
$K^-K^+\pi^-\nu_\tau$ and $K^-K^+K^-\nu_\tau$,
we evaluate the branching ratios (BR) by solving the relation
of the number of events
using a matrix of efficiencies and fake rates, given in 
Table~\ref{tbl:EfficiencyTable}.
The correlation matrix and the number of other BG events are evaluated by
using MC samples.
\begin{table}[t]
\begin{center}
\begin{tabular}{l|c|c|c|c}
\hline
 & \multicolumn{4}{c}{Generated decay mode} \\
\cline{2-5} 
Reconstructed & $\tau\to$      & $\tau\to$    & $\tau\to$    & $\tau\to$ \\
decay mode    & $\pi\pi\pi\nu$ & $K\pi\pi\nu$ & $KK\pi\nu$   & $KKK\nu$  \\
\hline
$\pi\pi\pi\nu$ & 0.23          & 0.076        & 0.023   & $7.3\times 10^{-3}$ \\
$K\pi\pi\nu$   & 0.012         & 0.172        & 0.049        & 0.023     \\
$KK\pi\nu$  & $4.0\times 10^{-5}$  & $4.7\times 10^{-3}$ & 0.129  & 0.060     \\
$KKK\nu$    & $2.8\times 10^{-7}$  & $1.4\times10^{-5}$ & $2.8\times 10^{-3}$ & 0.094 \\
\hline
\end{tabular}
\end{center}
\caption{Summary of the efficiencies and fake rates.}
\label{tbl:EfficiencyTable}
\end{table}

With the normalization correction by leptonic $\tau$ decays
and a systematic error evaluation,
we obtain a preliminary result of BR for
$\tau^- \to K^- \pi^+ \pi^- \nu_\tau$
as ${\cal B}=(3.25 \pm 0.02({\rm stat.}) ^{+0.16}_{-0.15}({\rm sys.}))\times 10^{-3}$,
together with 
${\cal B}(\tau^- \to K^- K^+ \pi^- \nu_\tau)=
(1.53 \pm 0.01({\rm stat.}) \pm 0.06({\rm sys.}))\times 10^{-3}$ and
${\cal B}(\tau^- \to K^- K^+ K^- \nu_\tau)=
(2.60 \pm 0.23({\rm stat.}) \pm 0.10({\rm sys.}))\times 10^{-5}$.
The dominant systematic uncertainty is an estimation of the track-finding
efficiency, 3.2\%.
Our result is consistent with the world average in PDG2006~\cite{PDG2006}.

\section{Study of $\tau^- \to K^*(892)^0 K^- \nu_\tau$}

\begin{figure}[t]
\centering
\includegraphics[width=12cm]{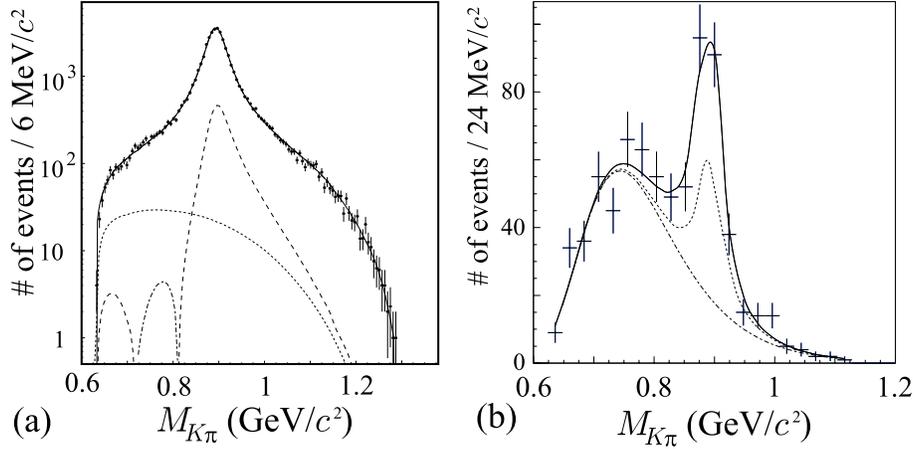}
\caption{Invariant mass distributions of
$K^+ \pi^-$ for the selection of
(a) $\tau^- \to K^{*0} K^- \nu_\tau$ and
(b) $\tau^- \to K^{*0} K^- \pi^0 \nu_\tau$.
The solid curve shows the fitted result.
In figure (a), the dashed, dot-dashed and dotted curves indicate
the spectra of the interference term, the inverse of
interference term and the non-resonant term, respectively.
In figure (b), the dotted and dash-dotted curves indicate
$K^{*0} K^- \nu_\tau$ contamination and the non-$K^{*0}$ BG.
}
\label{fig:mkstar}
\end{figure}

We study $\tau^- \to K^{*0} K^- \nu_\tau$ with data of 545fb$^{-1}$
by requiring the decay $K^{*0} \to K^+ \pi^-$ exclusively.
The signal yield is evaluated using the invariant mass distribution of
$K^+\pi^-$ with an additional selection of 
the $K^{*0}$ momentum, $p_{K^{*0}}^{CM}>1.5$GeV/$c$,
and the $K^{*0}K^-$ system momentum, $p_{K^{*0}K}^{CM}>3.5$GeV/$c$,
in the center-of-mass (CM) system.
After all selections, $5.1 \times 10^4$ events remain, as shown in
Fig.~\ref{fig:mkstar}(a). A clear $K^*(892)^0$ peak can be seen.

The dominant BG is from other $\tau$ decay modes.
The largest component, composing $\sim80$\% of the BG,
arises from $\tau^- \to K^- \pi^+ \pi^- n\pi^0 \nu_\tau$ and
$\tau^- \to \pi^- \pi^+ \pi^- n\pi^0 \nu_\tau$
through kaon mis-identification and an undetected $\pi^0$.
We evaluate the contamination in the $K^{*0}K^-\nu_\tau$ samples,
from data samples by replacing the kaon selection by the pion, 
and taking into account the PID fake rate.
Another $\sim20$\% of BG is from $\tau^- \to \phi K^- \nu_\tau$,
$K^-K^+\pi^-\pi^0\nu_\tau$ (excluding $K^{*0}$), $\phi \pi^- \nu_\tau$
and non-resonant $K^-K^+\pi^-\nu_\tau$, and is estimated by MC.

A possible peaking BG is $\tau^- \to K^*(892)^0 K^- \pi^0 \nu_\tau$;
however, this has not been measured yet. 
In order to measure the BR, 
${\cal B}(\tau^- \to K^*(892)^0 K^- \pi^0 \nu_\tau)$,
we apply the same criteria as that for the $K^{*0}K^- \nu_\tau$,
but with additional requirements of $\pi^0$.
Figure~\ref{fig:mkstar}(b) shows the $K^+\pi^-$ mass distribution
for $\tau^- \to K^{*0} K^- \pi^0 \nu_\tau$.
The large peaking BG is from $\tau^- \to K^{*0} K^- \nu_\tau$
with fake photons due to a hadronic shower in the calorimeter.
By applying a fit with the Breit-Wigner (BW) function for the $K^*(892)^0$
yield and the landau function for combinatorial BG,
we obtain a signal yield of $129\pm25$ events
after subtracting the peaking BG's of $114\pm7$ events.
The signal efficiency is 0.54\%.
As a result, we obtain the first measurement of the BR, 
${\cal B}(\tau^- \to K^*(892)^0 K^- \pi^0 \nu_\tau)
=(2.39\pm0.46\pm0.26) \times 10^{-5}$.

After the BG evaluation, we perform a fit to the $K^+\pi^-$ mass
distribution, with the following function:
$N_{\rm data}(M_{K\pi}) = 
F(M_{K\pi}) |\alpha A_{BW}(M_{K\pi}) + \beta A_{NR}(M_{K\pi})e^{i\phi}|^2
+ N_{\rm incoh}(M_{K\pi})$,
where $M_{K\pi}$ is the invariant mass of $K^+\pi^-$,
$F(M_{K\pi})$ is the phase space and kinematical factor,
$A_{BW}(M_{K\pi})$ is the BW function for $K^*(892)^0$,
considering the spin, parity and damping factor discussed 
in the LASS~\cite{LASS} and FOCUS~\cite{FOCUS}.
The term $A_{NR}(M_{K\pi})$ is the non-resonant $K^-K^+\pi^- \nu_\tau$ 
scalar amplitude, 
$A_{NR}(M_{K\pi})=(M_{K\pi}/q) \sin \delta_{\rm LASS} e^{i \delta_{\rm LASS}}$,
where $q$ is the momentum in the $K^+\pi^-$ center of mass
and $\delta_{\rm LASS}$ is the phase-shift,
given by LASS~\cite{LASS} and FOCUS~\cite{FOCUS}.
The term $N_{\rm incoh}(M_{K\pi})$ is the incoherent continuum and peaking BG's.
The free parameters in the $\chi^2$-fits are $\alpha$, $\beta$,
the $K^*(892)^0$ mass ($M_{K^*(892)^0}$), width (${\mit\Gamma}_{K^*(892)^0}$),
and the relative phase, $\phi$.

We obtain the fit result as shown in Fig.~\ref{fig:mkstar}(a)
and the BR of
${\cal B}(\tau^-\to K^*(892)^0 K^- \nu_\tau)=(1.56\pm0.02\pm0.09)\times10^{-3}$
with a non-resonant $K^-K^+\pi^-\nu_\tau$ BR of 
$(5.76\pm0.59\pm2.04)\times10^{-5}$.
The dominant systematic uncertainty is the errors of the track-finding
and the PID efficiencies.
We have measured ${\cal B}(\tau^- \to K^{*0} K^- \nu_\tau)$ in 
5.9\% precision, which is the most precise measurement ever obtained.
The non-resonant $K^- K^+ \pi^- \nu_\tau$ contribution is $(5.58\pm0.57)$\%,
which agrees well with the FOCUS result of $(5.30\pm0.74^{+0.99}_{-0.96})$\%.
The mass and width of $K^*(892)^0$ are obtained as
$M_{K^*(892)^0} = 895.10\pm0.27\pm0.31$~MeV/$c^2$ and
${\mit\Gamma}_{K^*(892)^0} = 47.23\pm0.49\pm0.79$~MeV.
Those are the first results using $\tau$ decay,
and are consistent with the result of FOCUS experiment~\cite{FOCUS}.

\section{Summary}

We have studied hadronic tau decays to the final state with three
charged hadrons using $\sim5\times 10^8$ $\tau$-pairs.
We evaluate the BR of $\tau^- \to K^- \pi^+ \pi^- \nu_\tau$,
$K^- K^+ \pi^- \nu_\tau$ and
$K^- K^+ K^- \nu_\tau$, simultaneously, in order to reduce the
systematic uncertainty of the cross-feed BG's, and obtain 
the preliminary result as follows:
\begin{eqnarray}
 {\cal B}(\tau^- \to K^- \pi^+ \pi^- \nu_\tau) &=& 
 (3.25\pm0.02^{+0.16}_{-0.15})\times 10^{-3}, \\
 {\cal B}(\tau^- \to K^- K^+ \pi^- \nu_\tau) &=& 
 (1.53\pm0.01\pm0.06)\times 10^{-3}, \\
 {\cal B}(\tau^- \to K^- K^+ K^- \nu_\tau) &=& 
 (2.60\pm0.23\pm0.10)\times 10^{-5}.
\end{eqnarray}
These results are consistent with the PDG average~\cite{PDG2006}.

We study $\tau^- \to K^- K^*(892)^0 \nu_\tau (\to K^- K^+ \pi^- \nu_\tau)$
and perform a fit to the $K^*(892)^0$ mass distribution by considering the
interference between the BW and the non-resonant amplitudes.
We also measure the BR of $\tau^- \to K^- K^{*0} \pi^0 \nu_\tau$.
The following BR's, mass and width of $K^*(892)^0$ are obtained:
\begin{eqnarray}
 {\cal B}(\tau^-\to K^*(892)^0 K^- \nu_\tau)&=&
  (1.56\pm0.02\pm0.09)\times10^{-3}, \\
 {\cal B}(\tau^-\to K^*(892)^0 K^- \pi^0 \nu_\tau)&=&
  (2.39\pm0.46\pm0.26) \times 10^{-5}, \\
 M_{K^*(892)^0} &=& 895.10\pm0.27\pm0.31 ~~{\rm MeV/}c^2, \\
 {\mit\Gamma}_{K^*(892)^0} &=& 47.23\pm0.49\pm0.79 ~~{\rm MeV}.
\end{eqnarray}
The high statistical data enable us to measure BR precisely as well as
the $K^*(892)^0$ mass and width for the first time using $\tau$ decays.

\begin{acknowledgments}
We thank the KEKB group for the excellent operation of the
accelerator, the KEK cryogenics group for the efficient
operation of the solenoid, and the KEK computer group and
the National Institute of Informatics for valuable computing
and SINET3 network support. We acknowledge support from
the Ministry of Education, Culture, Sports, Science, and
Technology of Japan and the Japan Society for the Promotion
of Science; the Australian Research Council and the
Australian Department of Education, Science and Training;
the National Natural Science Foundation of China under
contract No.~10575109 and 10775142; the Department of
Science and Technology of India; 
the BK21 program of the Ministry of Education of Korea, 
the CHEP src program and Basic Research program (grant 
No. R01-2005-000-10089-0, R01-2008-000-10477-0) of the 
Korea Science and Engineering Foundation;
the Polish State Committee for Scientific Research; 
the Ministry of Education and Science of the Russian
Federation and the Russian Federal Agency for Atomic Energy;
the Slovenian Research Agency;  the Swiss
National Science Foundation; the National Science Council
and the Ministry of Education of Taiwan; and the U.S.\
Department of Energy.
This work is supported by a Grant-in-Aid for Science Research on Priority Area 
(Mass Origin and Supersymmetry Physics, New Development of Flavor Physics)
from the Ministry of Education, Culture, Sports, Science and Technology 
of Japan.
\end{acknowledgments}

\end{document}